\documentclass[aps,pra,twocolumn,showpacs,superscriptaddress]{revtex4-1}
\usepackage{amssymb}
\usepackage{amsmath}
\usepackage{here}
\usepackage{soul}

\usepackage[pdftex]{graphicx,color}
\usepackage{hyperref}
\hypersetup{
   colorlinks=true,        
   linkcolor=blue,          
   citecolor=blue,        
  filecolor=magenta,      
   urlcolor=blue           
}

\begin{document}
\title{Magnetoelastic study on the frustrated quasi-one-dimensional spin-1/2 magnet LiCuVO$_4$}

\author{A. Miyata}
\affiliation{Hochfeld-Magnetlabor Dresden (HLD-EMFL) and W\"urzburg-Dresden Cluster of Excellence ct.qmat, Helmholtz-Zentrum Dresden-Rossendorf, 01328 Dresden, Germany}

\author{T. Hikihara}
\affiliation{Faculty of Science and Technology, Gunma University, Kiryu, Gunma 376-8515, Japan}

\author{S. Furukawa}
\affiliation{Department of Physics, Keio University, 3-14-1 Hiyoshi, Kohoku-ku, Yokohama, Kanagawa 223-8522, Japan}

\author{R. K. Kremer}
\affiliation{Max-Planck-Institut f\"ur Festk\"orperforschung, 70569 Stuttgart, Germany}

\author{S. Zherlitsyn}
\affiliation{Hochfeld-Magnetlabor Dresden (HLD-EMFL) and W\"urzburg-Dresden Cluster of Excellence ct.qmat, Helmholtz-Zentrum Dresden-Rossendorf, 01328 Dresden, Germany}

\author{J. Wosnitza}
\affiliation{Hochfeld-Magnetlabor Dresden (HLD-EMFL) and W\"urzburg-Dresden Cluster of Excellence ct.qmat, Helmholtz-Zentrum Dresden-Rossendorf, 01328 Dresden, Germany}
\affiliation{Institut f\"ur Festk\"orper- und Materialphysik, TU Dresden, 01069 Dresden, Germany}

\date{}

\begin{abstract}
We investigated the magnetoelastic properties of the quasi-one-dimensional spin-1/2 frustrated magnet LiCuVO$_4$. Longitudinal-magnetostriction experiments were performed at 1.5 K in high magnetic fields of up to 60 T applied along the $b$ axis, i.e., the  
spin-chain direction. 
The magnetostriction data qualitatively resemble the magnetization results, and saturate at $H_{\text{sat}} \approx 54$ T, with a relative change in sample length of $\Delta L/L \approx 1.8\times10^{-4}$. 
Remarkably, both the magnetostriction and the magnetization evolve gradually between $H_{\text{c3}} \approx 48$ T and $H_{\text{sat}}$, 
indicating that the two quantities consistently detect the spin-nematic phase just below the saturation. 
Numerical analyses for a weakly coupled spin-chain model reveal that the observed magnetostriction can overall be understood within an exchange-striction mechanism. 
Small deviations found may indicate nontrivial changes in local correlations associated with the field-induced phase transitions.
\end{abstract}

\maketitle
\section{Introduction}

One-dimensional (1D) 
quantum-spin magnets with competing interactions 
have been intensively studied for decades to search for unconventional quantum 
phases 
that result 
from the interplay between strong quantum fluctuations and magnetic frustration. 
As
1D systems generally 
allow 
simpler theoretical treatment compared with higher-dimensional ones, 
they provide an ideal platform for studying 
quantum phases 
through a close comparison between
experimental and theoretical approaches. One example is the spin-nematic phase, a magnetic analogue of the nematic liquid-crystal phase, where the spin-rotational symmetry is spontaneously broken while translational and time-reversal symmetries are preserved.  
This new type of quantum phase has been discussed in a variety of 1D spin models, 
including spin-1/2 \cite{Kec07,Hik08, Sud09, Bal16} and spin-$1$ \cite{Lau06, Har06} chains.
In the spin-1/2 case, 
magnon bound states (bimagnons) arise from a competition between ferromagnetic nearest-neighbour $J_1<0$ and antiferromagnetic next-nearest-neighbour $J_2>0$ interactions, and play a crucial role in the formation of the spin-nematic phase. Upon condensation, bimagnons form a Tomonaga-Luttinger liquid (TLL), which shows dominant spin density wave (SDW) or nematic 
quasi-long-range correlations at low and high fields, respectively. 

One well-studied spin-nematic candidate is the orthorhombic inverse spinel LiCuVO$_4$, where magnetic Cu$^{2+}$ ions form spin-1/2 chains with two dominant exchange interactions, $J_1$ and $J_2$, along the crystallographic $b$ axis~\cite{Pro04, End05, Koo11}.  
Weak interchain interactions (such as $J_5<0$) lead 
to a three-dimensional (3D) 
magnetic 
long-range order below $T_\text{c}\approx $ 2.3 K with an incommensurate planar spiral structure lying in the $ab$ plane \cite{Gib04,End10}. This spiral structure, which induces ferroelectricity through magnetoelectric coupling, 
has spurred
additional interest 
in 
LiCuVO$_4$ in terms of multiferroicity~\cite{Nai07,Sch08,Mou11,Ruf19}.

When applying magnetic fields, a phase transition to an incommensurate collinear 
SDW 
phase takes place at  $H_\text{c2}\approx 7.5$ T~\cite{But10, But12, Mas11}. 
Such a SDW phase is expected to appear when the SDW correlations of a bimagnon TLL are stabilized by interchain interactions \cite{Sat13,Sta14}. 
This picture has been corroborated 
by neutron scattering \cite{Mou12}, nuclear magnetic resonance (NMR) \cite{But14}, and spin Seebeck effect \cite{Hir19} experiments. 
With further increasing magnetic field, 
the SDW correlations are weakened in favor of the nematic correlations, 
and as a result, a 3D nematic long-range order 
is stabilized just below the saturation field, as theoretically discussed in Refs.~\cite{Sat13,Sta14, Zhi10, Ued09}. 
Remarkably, high-field magnetization~\cite{Svi11}, NMR~\cite{Orl17}, magnetocaloric effect (MCE), and ultrasound~\cite{Gen19} measurements have reported 
evidence of this 3D spin-nematic state between $H_{\text{c3}}$ and $H_{\text{sat}}$, which are about 40 and 45 T for the magnetic field applied along 
the 
$c$ axis, and about 48 and 53 T along 
the 
$a$ and $b$ axes.

While 
a large number of experiments have been performed on LiCuVO$_4$, there have been only few studies on its magnetoelastic properties.
Mourigal \textit{et al}.\ have proposed that the magnetostriction is negligible in LiCuVO$_4$ 
based on specific heat and neutron scattering experiments, 
and that the multiferroicity arises from a purely electronic spin-supercurrent mechanism~\cite{Mou11}. 
In contrast, recent thermal expansion and magnetostriction experiments in low magnetic fields of up to 9 T 
by Grams \textit{et al}.\ 
revealed a sizable magnetoelastic coupling in LiCuVO$_4$~\cite{Gra19}.

In this paper, we 
report 
longitudinal magnetostriction measurements of LiCuVO$_4$ at 1.5 K in high magnetic fields of up to 60 T applied along the $b$ axis, i.e., the 1D spin-chain direction. 
The observed magnetostriction is as large as $\Delta L/L \approx 1.8\times10^{-4}$ at high fields, 
indicating that LiCuVO$_4$ has  
a
sizable magnetoelastic coupling as 
reported previously \cite{Gen19, Gra19}.
The magnetostriction data qualitatively resemble the magnetization reported in Ref.~\cite{Orl17}. 
In particular, both quantities evolve gradually between $H_{\text{c3}}$ and $H_{\text{sat}}$, 
indicating that the magnetostriction can also probe 
the 3D spin-nematic state appearing just below the saturation. 
Furthermore, we analyze the magnetostriction  
data
of LiCuVO$_4$ within an exchange-striction model with exchange interactions modified linearly by the distance between the involved spins. 
The density matrix renormalization group (DMRG) and exact diagonalization (ED) 
analyses for 
the $J_1$-$J_2$(-$J_5$) spin-chain model 
show 
an overall 
agreement with our experimental results.  
Small deviations found may indicate nontrivial changes in local correlations associated with field-induced phase transitions.
A more refined treatment of the interchain coupling or introduction of additional interactions, such as a Dzyaloshinskii-Moriya (DM) term, 
is 
needed 
to explain the magnetoelastic properties 
of
LiCuVO$_4$ in general detail. 

\section{Methods}\label{sec:methods}

Longitudinal-magnetostriction measurements were performed by the optical fiber Bragg grating (FBG) method~\cite{Daou10} in pulsed magnetic fields of up to 60 T with a whole pulse duration of 25 ms at the HLD-EMFL in Dresden. The relative length change, 
 $\Delta L/L$, was obtained from the Bragg wavelength shift of the FBG.  A high-quality LiCuVO$_4$ single crystal with a size of 1.9 $\times$ 2.0 $\times$ 0.65 mm$^3$ was used in this study, which is the same one previously  
used 
in the magnetocaloric effect and ultrasound experiments 
by Gen \textit{et al}.\ \cite{Gen19}.
Magnetic field was applied along the orthorhombic $b$ axis, which is 
the 
spin-chain direction (${\bf H} \parallel \Delta L/L \parallel b$ axis). 
Note that the magnetic field was applied along the $c$ axis in Ref.~\cite{Gen19}. 

\newcommand{\uv}{\textbf{u}}
\newcommand{\vv}{\textbf{v}}
Numerical analyses were conducted for the $J_1$-$J_2$-$J_5$ model that describes magnetic interactions in the $ab$ plane. 
We use the ratio $J_1:J_2:J_5 = -0.42:1:-0.11$ as obtained from neutron-scattering experiments reported in Ref.\ \cite{End05}. 
For the Land\'e $g$-factor, we take the value $g=2.095$ along the $b$ axis from Ref.\ \cite{Kru02}. 
We then set $J_2=4.0$ meV so that the theoretical saturation field is adjusted to 48 T, at which the experimental magnetization curve is the steepest; see Appendix \ref{App:mag}. 
This value is quite close to $J_2=$ 3.8 meV reported in Ref.~\cite{End05}. 
DMRG calculations were performed for the pure 1D $J_1$-$J_2$ model with up to $168$ spins (see Appendix \ref{App:DMRG} for details).
We took into account the interchain coupling $J_5$ in a mean-field manner following Ref.\ \cite{Sta10}.
As a complementary analysis, ED calculations based on TITPACK ver.\ 2 \cite{titpack} were performed directly for the $J_1$-$J_2$-$J_5$ model with up to 28 spins on the 2D plane. 
Both the DMRG and ED results show an overall agreement with the experimental magnetization, as shown 
in Appendix \ref{App:mag}. 

\section{Results and Discussion}

\subsection{Magnetostriction measurements and exchange-striction mechanism}

\begin{figure}[]
\centering
\includegraphics[angle=0,width=0.48\textwidth]{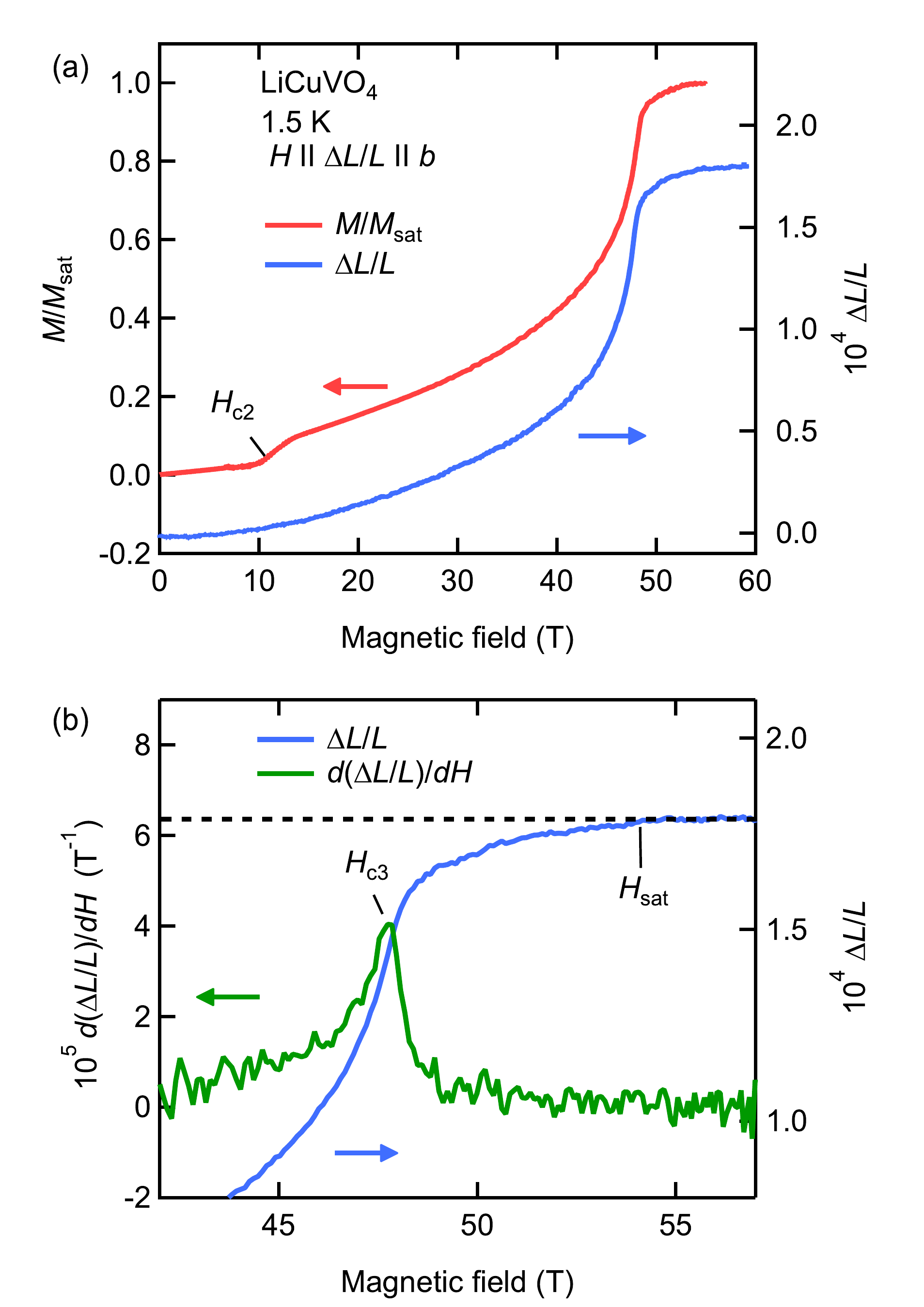}
\caption{(a) Longitudinal magnetostriction $\Delta L/L$ of LiCuVO$_4$ and  
the 
normalized magnetization $M$/$M_{\text{sat}}$ from Ref.~\cite{Orl17} as a function of magnetic field (${\bf H} \parallel \Delta L/L \parallel$ $b$ axis). 
(b) Longitudinal magnetostriction $\Delta L/L$ of LiCuVO$_4$ and its derivative in the vicinity of the saturation field. 
}
\label{fig:DL_H}
\end{figure}

Figure \ref{fig:DL_H}(a) shows the longitudinal magnetostriction, $\Delta L/L$, of LiCuVO$_4$ in magnetic fields of up to 60 T (${\bf H} \parallel \Delta L/L \parallel$ $b$ axis) with the normalized magnetization, $M/M_{\text{sat}}$, 
taken from Ref.~\cite{Orl17}. Note that the LiCuVO$_4$ sample used in magnetization measurements was from the same batch 
as that
used in  
the present 
magnetostriction study. 
The magnetostriction data qualitatively resemble 
the magnetization up to the saturation,  
except that no transition was detected at around 10 T (labelled as $H_{\text{c2}}$ in the magnetization). 
At saturation, the observed magnetostriction is 
as large as $\Delta L/L \approx 1.8 \times$10$^{-4}$, indicating a sizable magnetoelastic coupling in LiCuVO$_4$ as 
also reported recently~\cite{Gen19, Gra19}. 

To highlight the anomaly in high magnetic fields, 
the 
derivative of the longitudinal magnetostriction, $d(\Delta L/L)/dH$, is shown in Fig. \ref{fig:DL_H}(b).  
Notably, the magnetostriction still grows above 48 T ($H_{\text{c3}}$),  
at which 
a clear peak in $d(\Delta L/L)/dH$ is observed, and it saturates around 54 T ($H_{\text{sat}}$). Similar behavior has 
also been observed
in magnetization and NMR experiments, as reported previously in Refs.~\cite{Svi11,Orl17}; 
in these works, 
evidence for the existence of a 3D spin-nematic phase between $H_{\text{c3}}$ and $H_{\text{sat}}$
was given. 
The observed similarity indicates that the magnetostriction can consistently detect the 3D spin-nematic phase as well.

\begin{figure}[]
\centering
\includegraphics[angle=0,width=0.45\textwidth]{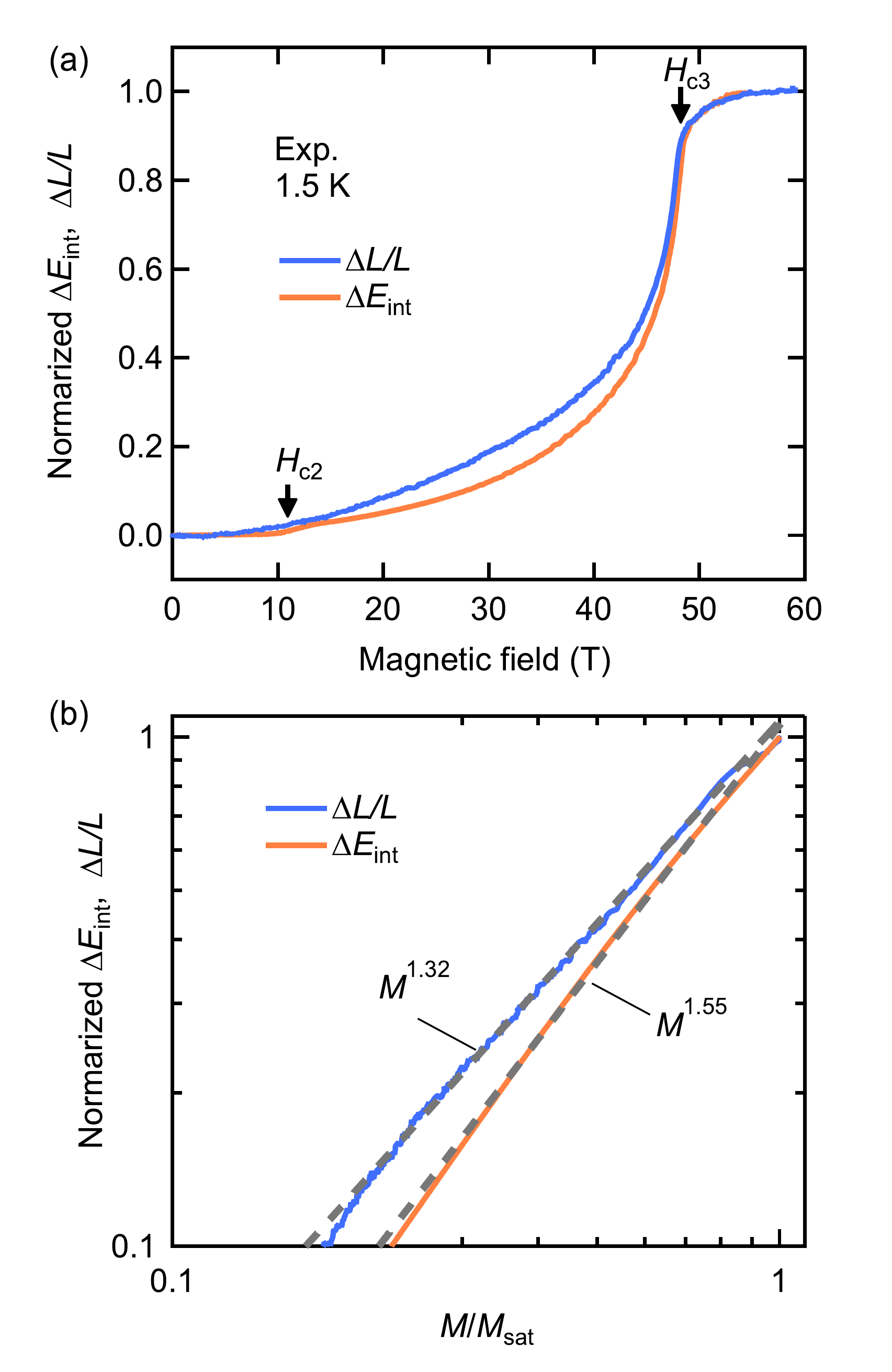}
\caption{
(a) Magnetostriction, $\Delta L/L$, and the spin-interaction energy, $\Delta E_{\text{int}}$, of LiCuVO$_4$ at 1.5 K versus magnetic field. 
Both quantities are normalized to 1 at saturation. 
The spin-interaction energy is obtained from the magnetization reported in Ref.\ \cite{Orl17} through the relation \eqref{eq:Eint_HM}. 
(b)  Logarithmic plot of 
$\Delta L/L$ and $\Delta E_{\text{int}}$ 
versus $M$. 
Dashed lines show power-law fits $\propto M^p$ with $p$ = 1.32 for $\Delta L/L$ and $p$ = 1.55 for $\Delta E_{\text{int}}$ (fit range is 0.4 $\leqq$ $M/M_\mathrm{sat}$ $\leqq$ 0.7). 
}
\label{fig:DL_Eint}
\end{figure}

To understand the microscopic origin of the magnetostriction, 
we adopt the following exchange-striction model for a 1D frustrated $J_1$-$J_2$ spin-chain system~\cite{Zap08, Ike19}
\begin{equation}\label{eq:H_es}
  \mathcal{H} _\text{es}=  \sum_{n=1,2}\sum_{j}  (J_\textit{n} - f_\textit{n}\epsilon){\bf S}_{\textit{j}}\cdot {\bf S}_{\textit{j}\text{+}\textit{n}} +\frac{N_\mathrm{1D}k'}{2}\epsilon^2  
  - g\mu_\mathrm{B} H \sum_{j}S_{\textit{j}}^{z} ,
\end{equation}
where $\epsilon = \Delta L/L$, 
$f_\textit{n}$ is the magnetoelastic coupling coefficient, $k'$ is the elastic constant per Cu$^{2+}$ ion, and $N_\mathrm{1D}$ is the number of Cu$^{2+}$ ions in the chain. 
For a fixed field $H$ or a fixed magnetization $M$, $\epsilon$ can be determined from minimizing the expectation value of $\mathcal{H} _\text{es}$. 
To compare with the experimental data, we set $\epsilon$ to zero for $M=0$, obtaining
\begin{equation}\label{eq:magnetostriction_corr}
 \epsilon(M) = \frac{1}{k'} \sum_{n=1,2} f_n \left( \overline{\langle{\bf S}_{j}\cdot {\bf S}_{\textit{j+n}}\rangle_M} - \overline{\langle{\bf S}_{j}\cdot {\bf S}_{\textit{j+n}}\rangle_{M=0}} \right), 
\end{equation}
where $\langle \cdots \rangle_M$ denotes the expectation value 
at 
the magnetization $M$ and $\overline{\cdots}$ indicates the spatial average (i.e., the average over $j$).
This expression indicates that the magnetostriction $\epsilon=\Delta L/L$ is a useful probe to detect local spin correlations $\overline{\langle{\bf S}_{j}\cdot {\bf S}_{j+n}\rangle_M}$, 
although there are unknown coefficients $f_n$. 
For the sake of simplicity, we have neglected the contribution of the interchain coupling $J_5$; by including it, the change in the local spin correlation on the $J_5$ bond will be added to Eq.\ \eqref{eq:magnetostriction_corr} with a coefficient $2f_5/k'$.

\begin{figure}[]
\centering
\includegraphics[angle=0,width=0.45\textwidth]{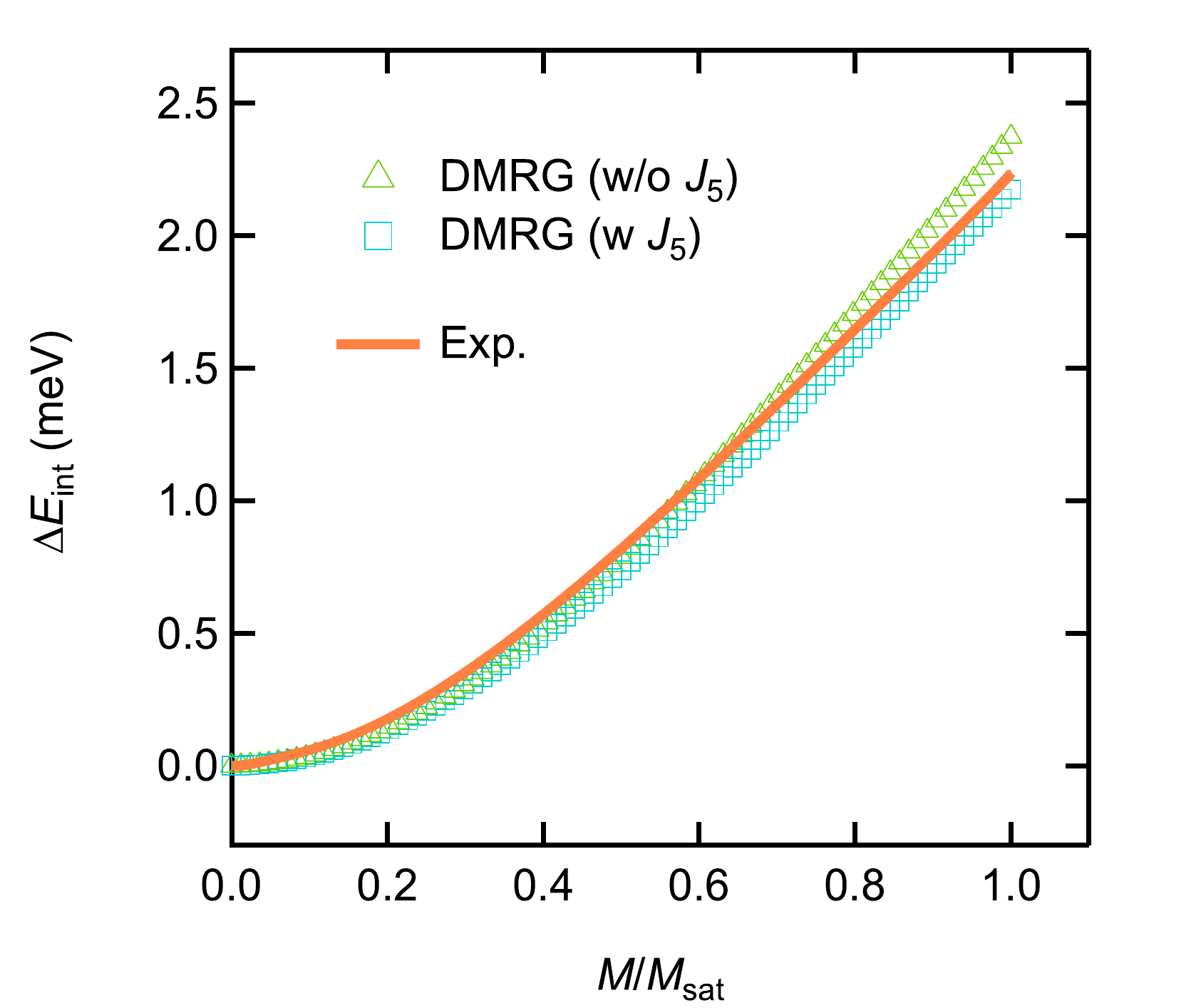}
\caption{
Comparison of the spin-interaction energy $\Delta E_{\text{int}}$ per spin extracted from the experimental data with the DMRG results with and without the interchain coupling $J_5$. 
}
\label{fig:Eint_M}
\end{figure}

As Eq.\ \eqref{eq:magnetostriction_corr} is similar to 
the expression 
of the spin-interaction energy $E_\mathrm{int}$ (i.e., the expectation value of the Heisenberg interaction part of the Hamiltonian), 
it is interesting to compare $E_\mathrm{int}$ with the measured $\Delta L/L$. 
Exploiting the thermodynamic relation $\frac{\partial E_\mathrm{int}}{\partial M}=H$, the spin-interaction energy relative to the zero-field case 
can be obtained from the magnetization data ($M$ versus $H$) via 
\begin{equation}\label{eq:Eint_HM}
\Delta E_{\text{int}}(M)\equiv E_{\text{int}}(M) - E_{\text{int}}(0) = \int_{0}^{M}H(M')dM'.
\end{equation}
$\Delta L/L$ and $\Delta E_{\text{int}}$ are shown in Fig.\ \ref{fig:DL_Eint}, where both quantities are normalized to 1 at saturation. 
As seen in Fig.\ \ref{fig:DL_Eint}(a), the two curves exhibit an overall similar field dependence. A small deviation is found in the intermediate SDW phase between $H_{\text{c2}}$ and $H_{\text{c3}}$. 
This 
indicates 
that the exchange-striction model, Eqs.\ \eqref{eq:H_es} and \eqref{eq:magnetostriction_corr}, is a good approximation to describe the magnetoelastic properties of LiCuVO$_4$.   
We note that both $\Delta L/L$ and $\Delta E_{\text{int}}$ show a kink around $H_\mathrm{c3}$; 
The kink in $\Delta E_{\text{int}}$ is consistent with the theoretical picture of Ref.\ \cite{Zhi10}.

\newcommand{\Sv}{\textbf{S}}
\newcommand{\qv}{\textbf{q}}
\newcommand{\rv}{\textbf{r}}

To analyze further the slightly different behavior of $\Delta L/L$ and $\Delta E_{\text{int}}$, we plot them as a function of $M/M_\mathrm{sat}$ in logarithmic scales in Fig.\ \ref{fig:DL_Eint}(b). 
Our motivation for this plot stems from the fact that the power-law relation $\Delta L/L\propto M^p$ has been observed in a variety of magnetic materials: with
$p=2$ in conventional antiferromagnets, $p=1$ in spin-dimer systems \cite{Jai12, Saw05}, 
and $p=1.3$ in the distorted kagome-lattice magnet volborthite, Cu$_3$V$_2$O$_7$(OH)$_2\cdot$2H$_2$O \cite{Ike19}. 
Here, $p=2$ can be understood from a classical canted antiferromagnetic order, in which $\Sv_i=S(\sin\theta\cos \qv\cdot\rv_i,\sin\theta\sin\qv\cdot\rv_i,\cos\theta)$, with $\theta$ and $\qv$ being the canting angle and the propagation vector, respectively. This gives
\begin{equation}
\begin{split}
 \Sv_i\cdot\Sv_j &=S^2\left(\sin^2\theta \cos \qv\cdot\rv_{ij}+\cos^2\theta \right) \\
 &=(S^2-m^2)\cos\qv\cdot\rv_{ij}+m^2,
\end{split}
\end{equation}
with $m=S_i^z=S\cos\theta$ and $\rv_{ij}=\rv_i-\rv_j$. 
$p=1$ in spin-dimer systems is also comprehensible from the fact that the magnetization $M$ and the change in the intradimer correlations $\langle \Sv_i\cdot\Sv_j\rangle_M$ are both proportional to the number of triplons. 
The unusual power $p=1.3$ in volborthite could, thus, be interpreted as a result of strong quantum fluctuations in a low-dimensional frustrated magnet. 
As shown in Fig.\ \ref{fig:DL_Eint}(b), our data fit well to the relations $\Delta L/L\propto M^{1.32}$ and $\Delta E_{\text{int}}\propto M^{1.55}$ in the SDW phase. 
We will use these power-law relations as a guide to compare the experimental and numerical results. 
The exponents $p=1.32$ and $1.55$ are significantly different from that of conventional antiferromagnets; the former agrees with that for volborthite. 
This indicates a crucial role of quantum fluctuations in LiCuVO$_4$ due to its quasi-1D nature.

\begin{figure}[]
\centering
\includegraphics[angle=0,width=0.45\textwidth]{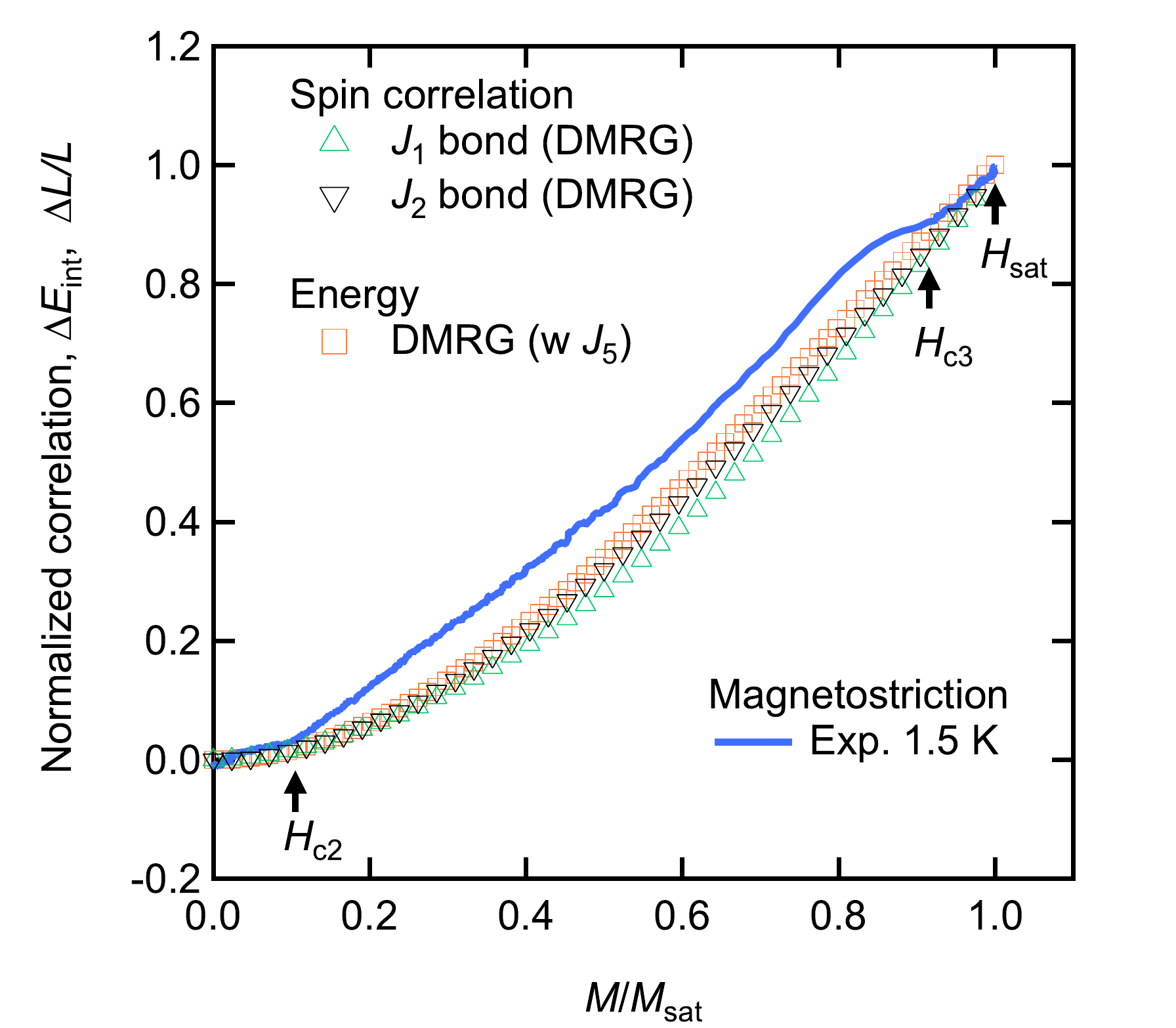}
\caption{
Comparison of the magnetostriction results 
with the DMRG analyses on the local spin correlations $\overline{\langle{\bf S}_{\textit{i}}\cdot {\bf S}_{\textit{j}}\rangle_M}$ on the $J_1$ and $J_2$ bonds 
and the spin interaction energy $\Delta E_{\text{int}}$. 
Here, all the quantities are normalized between 0 and 1, which correspond to the zero-field case and the saturation, respectively.
}
\label{fig:DL_corr_M}
\end{figure}

\subsection{Comparison with numerical results}

In Fig.\ \ref{fig:Eint_M}, we compare the spin-interaction energy $\Delta E_{\text{int}}$ per spin extracted from the experimental data with the DMRG results with and without the interchain coupling $J_5$. 
Here, the DMRG calculations were carried out for the pure 1D $J_1$-$J_2$ model, and
the effect of $J_5$ is taken into account in a mean-field manner by simply adding $2J_5(M/2M_{\text{sat}})^2$ to the energy per spin \cite{Sta10}. 
The DMRG results modified by $J_5$ show a very good agreement with the experiments even without normalization for the vertical axis. 
The agreement is expected since $\Delta E_{\text{int}}$ is related to the magnetization via Eq.\ \eqref{eq:Eint_HM}, 
and we have adjusted the value of $J_2$ in such a way that the DMRG result modified by $J_5$ 
can best reproduce the experimental magnetization (Appendix \ref{App:mag}). 

\begin{figure}[]
\centering
\includegraphics[angle=0,width=0.45\textwidth]{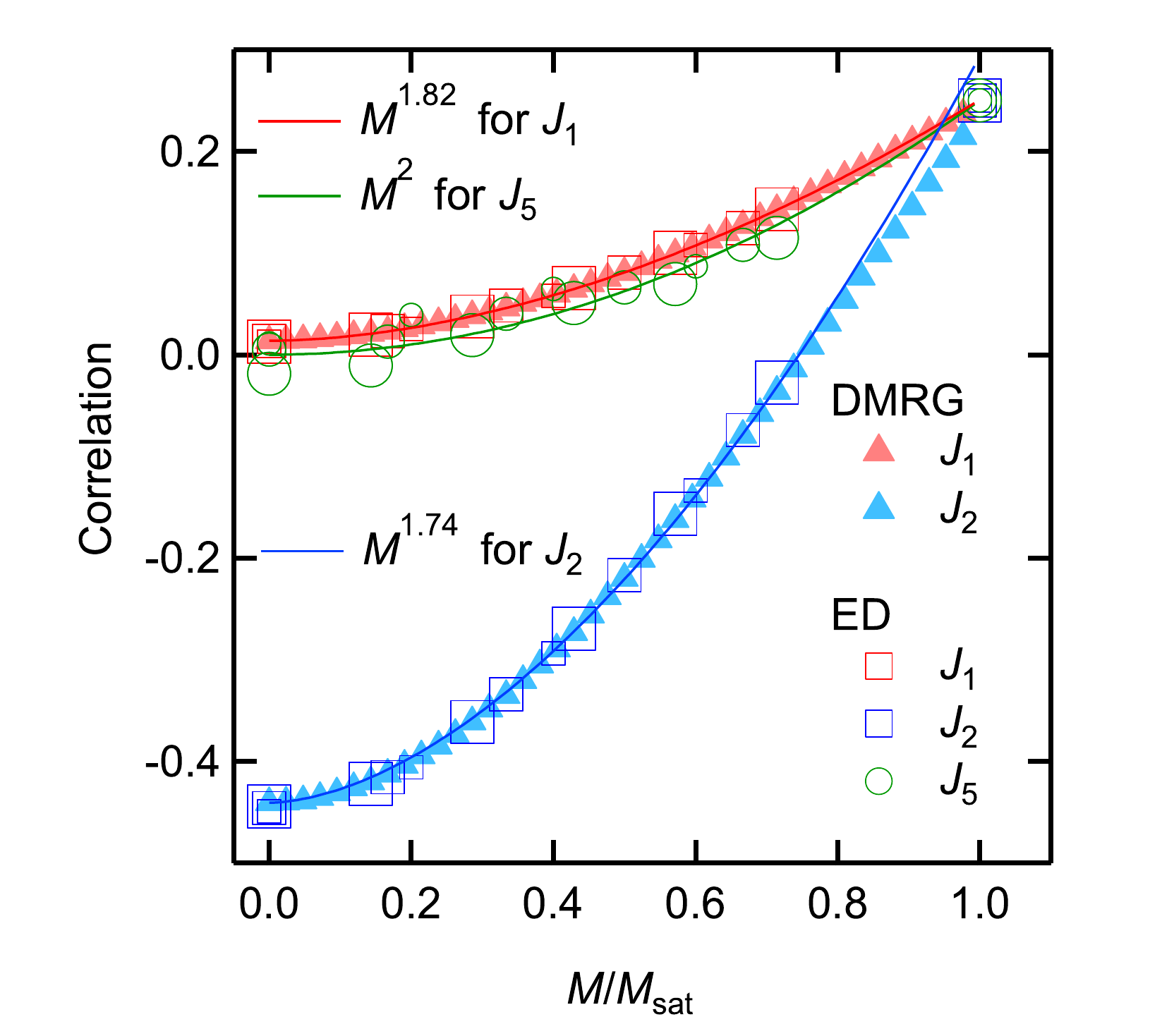}
\caption{Local spin correlations $\overline{\langle{\bf S}_{\textit{i}}\cdot {\bf S}_{\textit{j}}\rangle_M}$ on the $J_1$, $J_2$, and $J_5$ bonds calculated using DMRG and ED. 
Solid lines show 
the scaling [Eq. \eqref{eq:corr_scaling}] 
with $p$ =1.82, 1.74, and 2 for the $J_1$, $J_2$, and $J_5$ bonds, respectively \cite{note-pow-fit}.
In the ED results, symbols with small, medium, and large sizes correspond to the cases of 20, 24, and 28 spins, respectively [a screw boundary condition \cite{Nis09} 
with $(L_x,L_y)=(5,4)$, $(6,4)$, and $(7,4)$ as described in Appendix \ref{App:mag}].
}
\label{fig:corr_M}
\end{figure}

In Fig.\ \ref{fig:DL_corr_M}, we compare the experimental magnetostriction data
with the DMRG results of the local spin correlations $\overline{\langle{\bf S}_{\textit{i}}\cdot {\bf S}_{\textit{j}}\rangle_M}$ on the $J_1$ and $J_2$ bonds 
and the spin-interaction energy $\Delta E_{\text{int}}$
(see Appendix \ref{App:DMRG} for details of the DMRG analysis).
Here, the spatial average $\overline{\cdots}$ is taken over the concerned bonds $(i,j)$ for each interaction, and 
all the quantities are normalized between 0 and 1, which correspond to the zero-field case and the saturation, respectively. 
We also present the DMRG and ED results on the local spin correlations without normalization in Fig.\ \ref{fig:corr_M}. 
Notably, in Fig.\ \ref{fig:DL_corr_M}, the normalized spin correlations on the $J_1$ and $J_2$ bonds show quite similar behavior. 
This is remarkable as $J_1$ and $J_2$ are ferromagnetic and antiferromagnetic, respectively, and 
the correlations have opposite signs at low fields, as seen in Fig.\ \ref{fig:corr_M}. 
While we do not have a physical explanation for this behavior, it has the convenient consequence that
any linear combination of the two local correlations shows similar behavior if normalized. 
Indeed, in Fig.\ \ref{fig:DL_corr_M}, the DMRG data for the spin-interaction energy $\Delta E_{\text{int}}$ also behave similarly to the local spin correlations 
although it contains the small mean-field contribution of the interchain coupling $J_5$. 
In contrast, the experimental magnetostriction data, $\Delta L/L$, which is also expected to be described by the local spin correlations according to Eq.\ \eqref{eq:magnetostriction_corr}, 
show a clear deviation in the SDW phase between $H_{\text{c2}}$ and $H_{\text{c3}}$. 
This deviation cannot be explained even if we consider a linear combination of the spin correlations on $J_1$, $J_2$, and also $J_5$ bonds, as we explain in the following. 
The calculated data in Fig.\ \ref{fig:corr_M} are described well by the relation 
\begin{equation}\label{eq:corr_scaling}
 \overline{\langle{\bf S}_{\textit{i}}\cdot {\bf S}_{\textit{j}}\rangle_{M}}-\overline{\langle{\bf S}_{\textit{i}}\cdot {\bf S}_{\textit{j}}\rangle_{M=0}} \propto M^p,
\end{equation}
with $p=$ 1.82, 1.74, and 2 for the $J_1$, $J_2$, and $J_5$ bonds, respectively. 
Note that the relation with $p=2$ for the $J_5$ bond is automatically satisfied in the mean-field treatment of $J_5$, 
and it is confirmed well in the ED result for the 2D model. 
If we assume that the contribution of the $J_5$ bond is an order of magnitude smaller, 
a linear combination of the correlations (Fig.\ \ref{fig:corr_M}) cannot lead to the scaling of the magnetostriction $\Delta L/L\sim M^{1.32}$ in the SDW phase [Fig.\ \ref{fig:DL_Eint}(b)].  
Furthermore, such a linear combination cannot explain the kink-like features of the magnetostriction around the field-induced transition points $H_\mathrm{c2}$ and $H_\mathrm{c3}$ (Fig.\ \ref{fig:DL_corr_M}). 

The above consideration indicates a missing contribution in the magnetostriction. 
Such a contribution should be sensitive to the field-induced phase transitions as observed experimentally. 
A more refined treatment of the interchain coupling $J_5$ might be required. 
It would also be important to investigate the effects of additional terms such as exchange anisotropy and DM interactions in the spin Hamiltonian and in the exchange-striction mechanism described by Eqs.\ \eqref{eq:H_es} and \eqref{eq:magnetostriction_corr}.

\section{Summary}

Magnetostriction measurements of LiCuVO$_4$ in high magnetic fields up to 60 T applied along the $b$ axis 
show monotonous increase of  $\Delta L/L$ reaching a sizable value $\Delta L/L \approx 1.8\times10^{-4}$ at the saturation field $H_{\text{sat}}\approx 54$ T. 
Both the magnetostriction and the magnetization~\cite{Orl17} evolve in a similar way between $H_{\text{c3}}\approx 48$ T and $H_{\text{sat}}\approx 54$ T, 
which indicates that both quantities consistently detect the 3D spin-nematic phase just below saturation. 
Our results were discussed within the exchange-striction mechanism. 
The DMRG and ED analyses for 
the $J_1$-$J_2$(-$J_5$) 
spin-chain model show a good agreement with the experimental observations. 
A more refined treatment of the interchain coupling or additional interactions such as DM interactions would be needed 
to explain the magnetoelastic properties of LiCuVO$_4$ in more detail. 
As our results reveal the importance of the magnetoelastic coupling, it would be interesting to reconsider the contribution of the lattice to the multiferroic properties of LiCuVO$_4$, which is under debate \cite{Mou11,Gra19}.

\begin{acknowledgments}
We acknowledge support of the HLD at HZDR, member of the European Magnetic Field Laboratory (EMFL), the
DFG through SFB 1143 and the W\"{u}rzburg-Dresden Cluster of Excellence on
Complexity and Topology in Quantum Matter--$ct.qmat$ (EXC 2147, Project
No.\ 390858490), and the BMBF via DAAD
(Project ID 57457940).
T.H. was supported by JSPS KAKENHI Grant Numbers JP17H02931 and JP19K03664.
S.F. was supported by JSPS KAKENHI Grant Number JP18K03446.
\end{acknowledgments}


\appendix

\section{DMRG calculations}\label{App:DMRG}

\begin{figure}[]
\centering
\includegraphics[angle=0,width=0.4\textwidth]{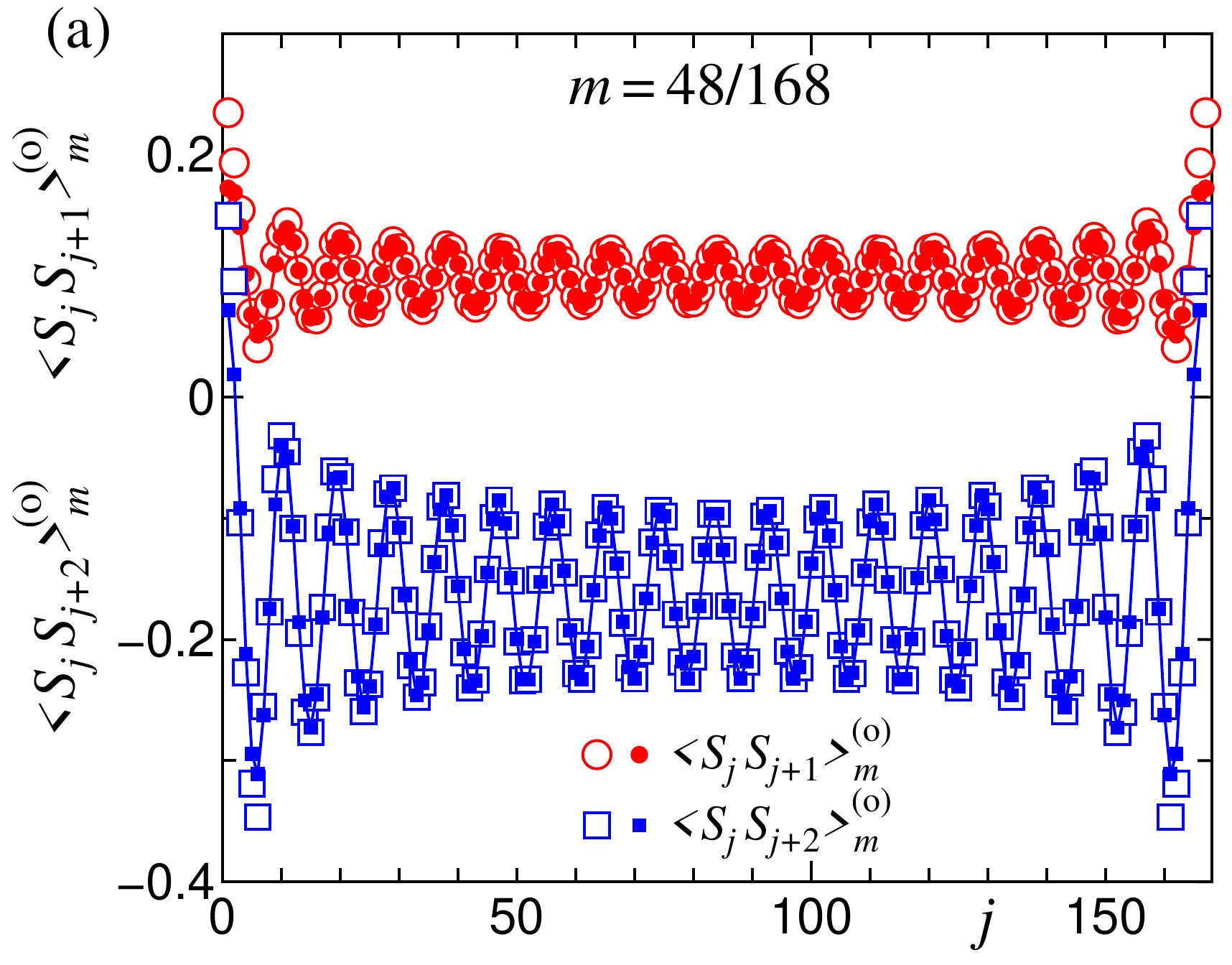}~~~~~~~~~~~\\
\includegraphics[angle=0,width=0.46\textwidth]{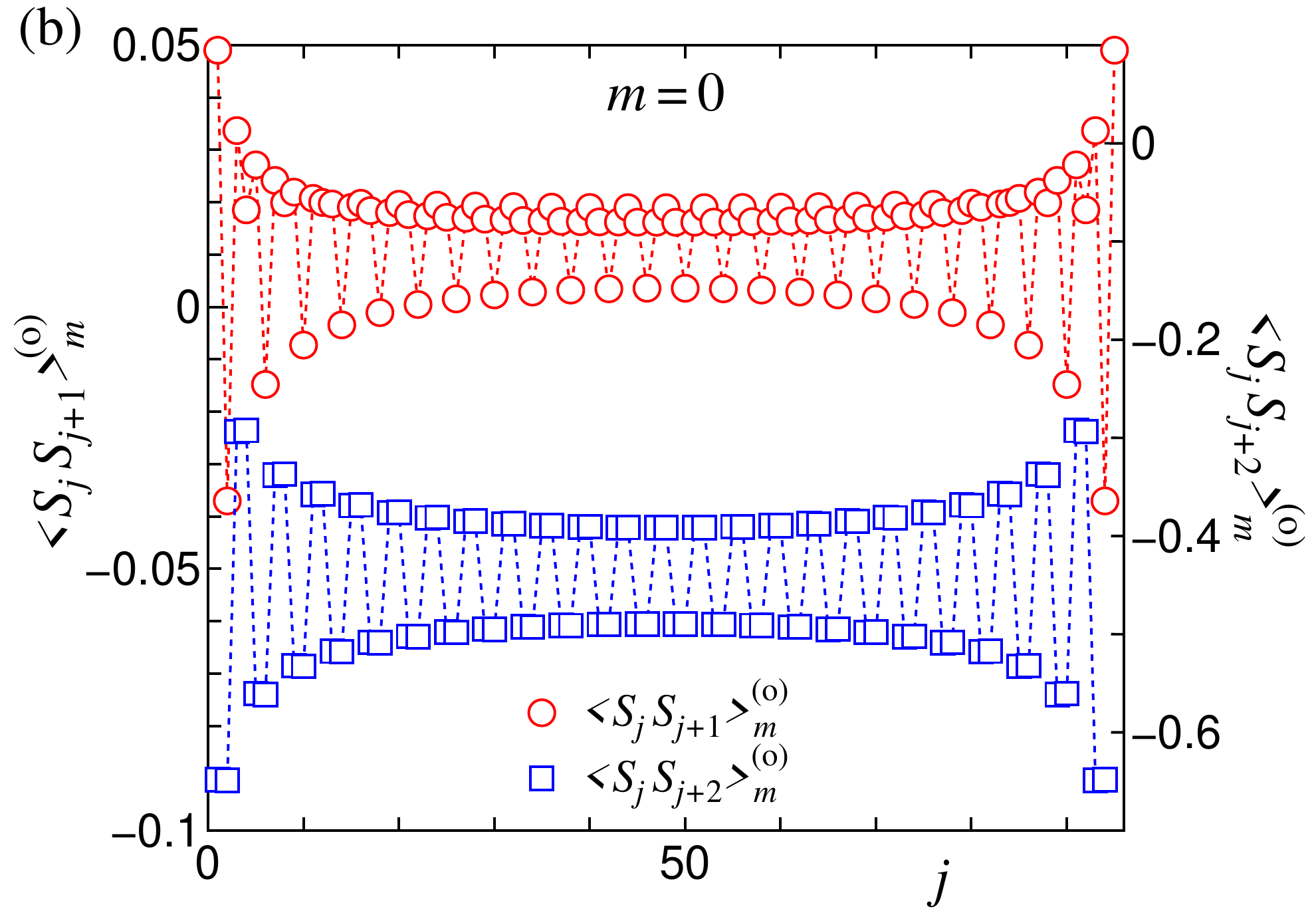}
\caption{
(a) Local spin correlations $\langle {\bf S}_j \cdot {\bf S}_{j+n}\rangle_m^{\rm (o)}~(n=1,2)$ for $N_\mathrm{1D}=168$ and $m=48/168$.
Open and solid symbols respectively represent the DMRG data and fits using
Eqs.\ (\ref{eq:cor1-form}) and (\ref{eq:cor2-form}).
Solid lines connecting the fits are guides to the eye.
(b) The DMRG data of $\langle {\bf S}_j \cdot {\bf S}_{j+n}\rangle_m^{\rm (o)}$ for $m=0$.
The data for $N_\mathrm{1D}=96$ are shown to demonstrate oscillations with four-site period.
We observed similar four-site period oscillations also for $N_\mathrm{1D}=120$ and $168$.
Dashed lines connecting the data points are guides to the eye.
}
\label{fig:local_cor}
\end{figure}

\begin{figure}[]
\centering
\includegraphics[angle=0,width=0.45\textwidth]{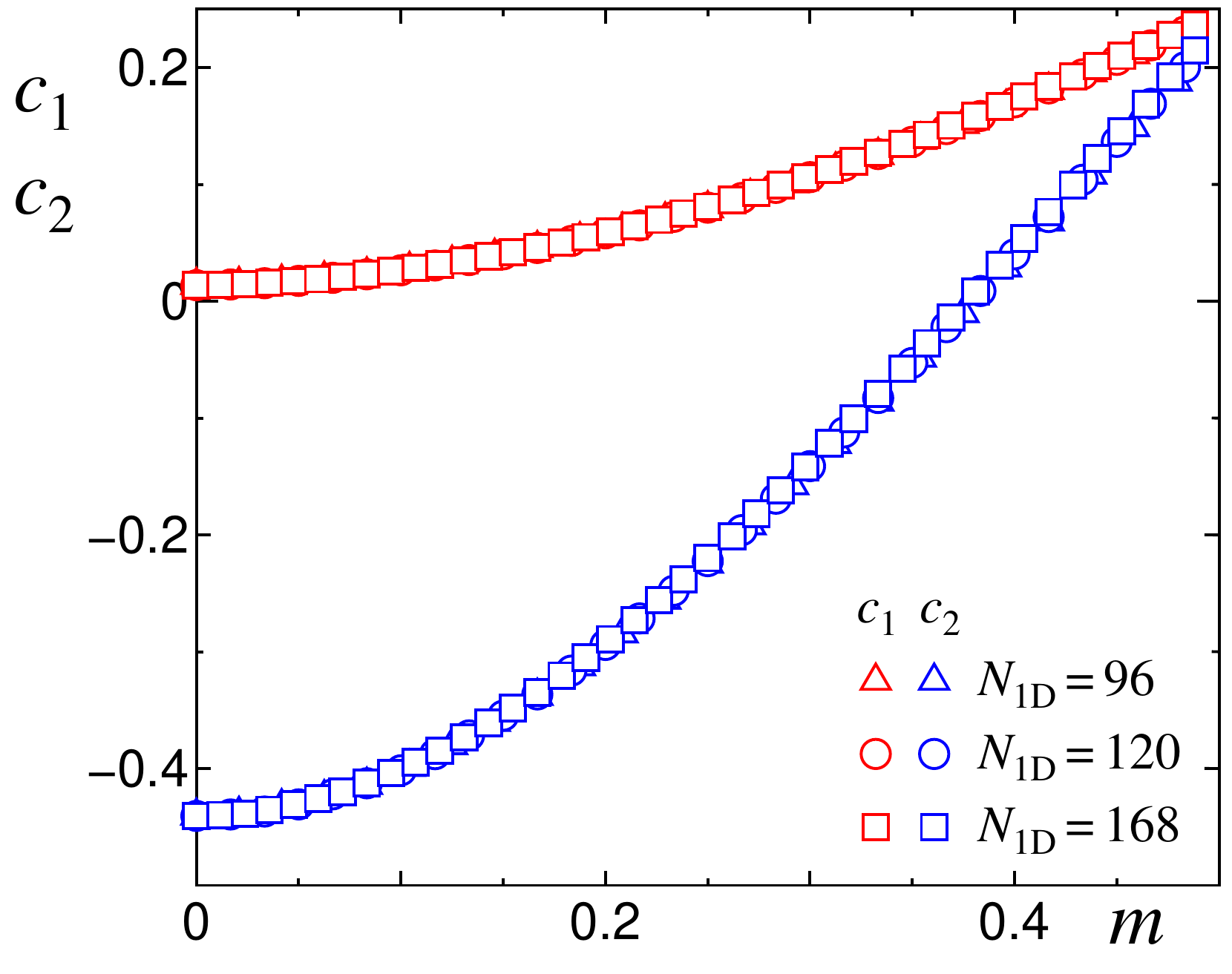}
\caption{
Uniform parts of the local spin correlations, $c_1$ and $c_2$, 
as functions of $m$. These were obtained from the fitting of the DMRG data of $\langle {\bf S}_j \cdot {\bf S}_{j+n}\rangle_m^{\rm (o)}$ for $N_\mathrm{1D}=96, 120,$ and $168$ with Eqs.\ (\ref{eq:cor1-form}) and (\ref{eq:cor2-form}) as shown in Fig.\ \ref{fig:local_cor}(a).
}
\label{fig:local_cor-m}
\end{figure}

We performed DMRG calculations for the  
pure 1D
$J_1$-$J_2$ Heisenberg model with the Hamiltonian given by
\begin{eqnarray}
\mathcal{H} _\text{1D} 
= J_1 \sum_j {\bf S}_j \cdot {\bf S}_{j+1}
+ J_2 \sum_j {\bf S}_j \cdot {\bf S}_{j+2}.
\label{eq:Ham_1D}
\end{eqnarray}
We set $J_1:J_2=-0.42:1$ as explained in Sec.\ \ref{sec:methods}.
The number of spins was up to $N_\mathrm{1D}=168$ and open boundary conditions were imposed.

Using DMRG, we calculated the lowest-energy state in each subspace characterized by the $z$ component of the total spin, $S^z_{\rm tot}=\sum_j S^z_j$.
We then computed the lowest energy $E_\mathrm{int}^\mathrm{1D}(m)$ and the local spin correlations $\langle {\bf S}_j \cdot {\bf S}_{j+n}\rangle_m^{\rm (o)}$ ($n=1,2$), where $m=S^z_{\rm tot}/N_\mathrm{1D}$ and $\langle \cdots \rangle_m^{\rm (o)}$ denotes the expectation value with respect to the lowest-energy state in the subspace with $S^z_{\rm tot}=N_\mathrm{1D} m$ in the system under the open boundary conditions.
Note that $m$ relates to the magnetization $M$ in the main text as 
$M=N g \mu_B m$, where $N$ is the number of Cu$^{2+}$ ions. 
The number of density matrix eigenstates kept in DMRG that is 
required for achieving a sufficient accuracy 
depends on $N_\mathrm{1D}$ and $S^z_{\rm tot}$.
In our calculation, we kept up to $600$ states in the most severe case and the truncation error (the sum of the density matrix weights of discarded states averaged over the last sweep) was, at most, $6 \times 10^{-7}$.
We, thereby, confirmed that the calculation was accurate enough for our argument.

The magnetization, which will be discussed in Appendix \ref{App:mag}, was
obtained from the data of $E_\mathrm{int}^\mathrm{1D}(m)$ by 
finding $m$ that minimizes $E_\mathrm{int}^\mathrm{1D}(m)-h N_\mathrm{1D}m$ for each $h=g\mu_\mathrm{B}H$.
We found that $S^z_{\rm tot}$ appearing in the magnetization curve 
was all even except for the single case of $S^z_{\rm tot}=1$.
This observation indicates the formation of bound magnon pairs for $S^z_{\rm tot} \ge 2$.
We note that this is consistent with the known result that the $J_1$-$J_2$ Heisenberg chain  
with $J_1 < 0$, $J_2 > 0$, and $J_1/J_2 \gtrsim -1$ exhibits the Haldane dimer phase for $m=0$ and the bimagnon TLL phase for $0 < m < 1/2$, while the range of the vector-chiral phase 
inbetween them, if any, is too narrow to be detected in the numerical calculation \cite{Hik08,Sud09,HMeisnerMK2009,FurukawaSOF2012}. 
We consider only the states with even $S^z_{\rm tot}$ in the following analysis of the local spin correlations.

In Fig.\ \ref{fig:local_cor}, we present the DMRG data of the local spin correlations $\langle {\bf S}_j \cdot {\bf S}_{j+n}\rangle_m^{\rm (o)}~(n=1,2)$.
Since the translational symmetry is broken by the open boundary conditions, the local spin correlations contain sizable contributions of boundary oscillation.
We then found that, as shown in Fig.\ \ref{fig:local_cor}(a), the local spin correlations for $S^z_{\rm tot}=N_\mathrm{1D} m \ge 2$ were reproduced by the formulas,
\begin{eqnarray}
\langle {\bf S}_j \cdot {\bf S}_{j+1}\rangle_m^{\rm (o)}
&=& c_1 + c'_1 \frac{(-1)^j \cos\left[Q(j+1/2)\right]}{\left[f(2j+1)\right]^K},
\label{eq:cor1-form} \\
\langle {\bf S}_j \cdot {\bf S}_{j+2}\rangle_m^{\rm (o)}
&=& c_2 - c'_2 \frac{(-1)^j \sin\left[Q(j+1)\right]}{\left[f(2j+2)\right]^K},
\label{eq:cor2-form}
\end{eqnarray}
with
\begin{eqnarray}
f(x)=\frac{2(N_\mathrm{1D}+1)}{\pi} \sin\left(\frac{\pi |x|}{2(N_\mathrm{1D}+1)}\right),
\label{eq:fx}
\end{eqnarray}
that were derived as the expressions of the local energy density of the bimagnon TLL under open boundary conditions \cite{HikiharaFL2017,Hik08,HikiharaF2004}. 
The wave number $Q$ relates to $m$ via the density of bimagnons $\rho$ as
\begin{eqnarray}
Q&=&\frac{2\pi N_\mathrm{1D}}{N_\mathrm{1D}+1} \left(\rho-\frac{1}{2}\right),
\\
\rho&=&\frac{1}{2}\left(\frac{1}{2}-m \right).
\end{eqnarray}
We performed the least-square fitting of the DMRG data of $\langle {\bf S}_j \cdot {\bf S}_{j+n}\rangle_m^{\rm (o)}$ ($n=1,2$) for $N_\mathrm{1D}/4+1 \le j \le 3N_\mathrm{1D}/4-n$ 
with Eqs.\ (\ref{eq:cor1-form}) and (\ref{eq:cor2-form}) by taking $c_n$, $c'_n$, and $K$ as fit parameters.
We, thereby, determined the uniform parts of the correlations, $c_1$ and $c_2$, and employed them  
as the estimates of the local spin correlations $\overline{\langle{\bf S}_{\textit{i}}\cdot {\bf S}_{\textit{j}}\rangle_{M}}$ on the $J_1$ and $J_2$ bonds discussed in the main text \cite{fitting-quality}. 

For $S^z_{\rm tot}=0$, where the system is in the Haldane-dimer phase \cite{FurukawaSOF2012}, the local spin correlations are not described by the formulas (\ref{eq:cor1-form}) and (\ref{eq:cor2-form}).
Instead, we found that the correlations oscillate with a four-site period as shown in Fig.\ \ref{fig:local_cor}(b).
We, hence, took the average of the correlations at four bonds around the center of the chain as the estimates of the uniform parts of the correlations.

Figure\ \ref{fig:local_cor-m} shows the uniform parts, $c_1$ and $c_2$, of the local spin correlations for $N_\mathrm{1D}=96, 120, 168$ as functions of $m$.
The data exhibit a smooth behavior and the dependence on the system size $N_\mathrm{1D}$ is negligibly small.
We, thus, used the data of $c_1$ and $c_2$ for $N_\mathrm{1D}=168$ as the estimates of the local spin correlations $\overline{\langle {\bf S}_i \cdot {\bf S}_j\rangle_M}$ on the $J_1$ and $J_2$ bonds  
in the thermodynamic limit.

\section{Magnetization}\label{App:mag}

\begin{figure}[]
\centering
\includegraphics[angle=0,width=0.45\textwidth]{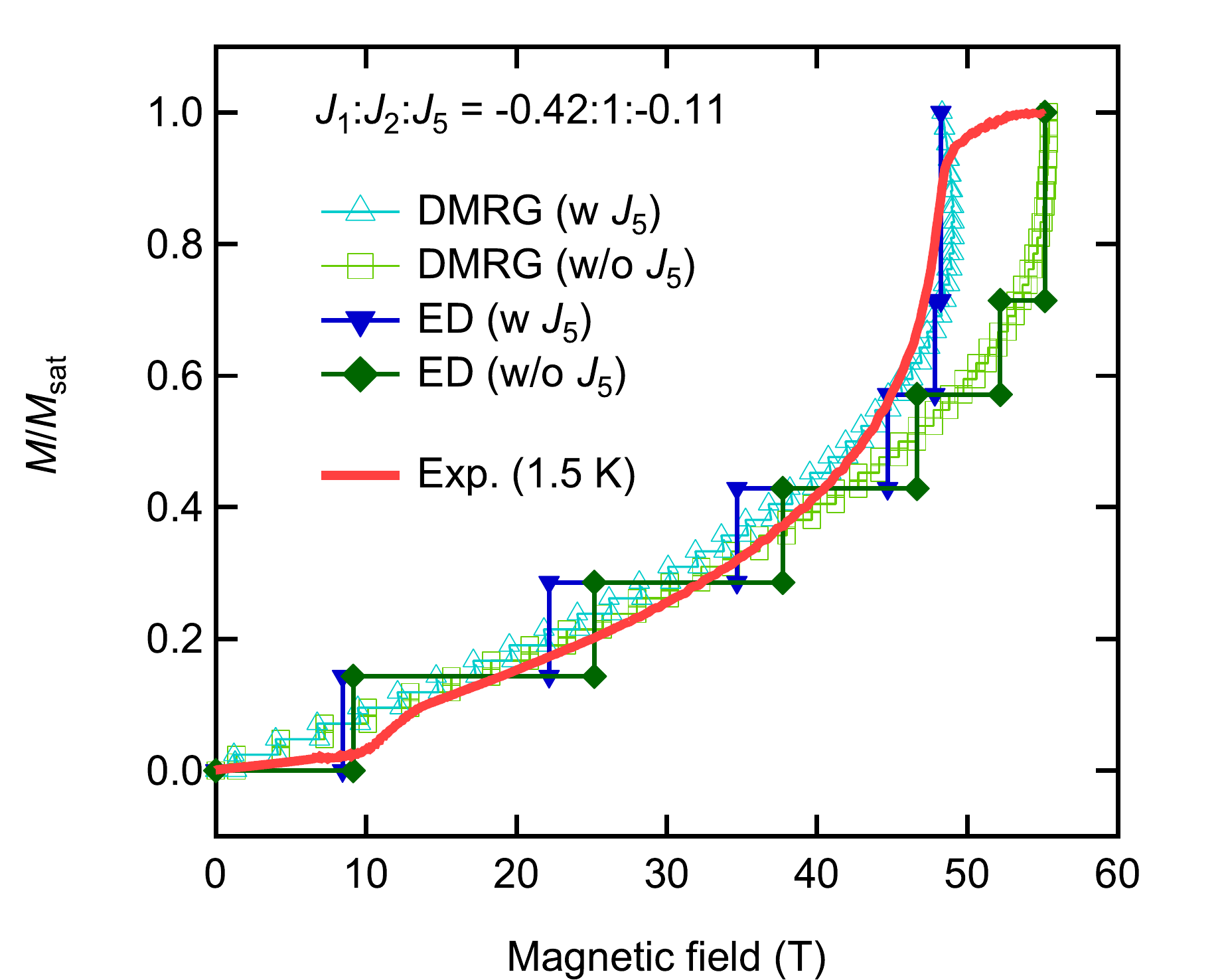}
\caption{
Comparison of the magnetization data of LiCuVO$_4$ from Ref.\ \cite{Orl17} with the DMRG and ED results for the $J_1$-$J_2$-$J_5$ spin-chain model. 
Numerical results are presented for the cases with and without the interchain coupling $J_5$ to highlight its effect. 
In the DMRG result, $J_5$ is treated in a mean-field manner via a shift of the magnetic field by $4J_5(M/2M_\mathrm{sat})/g\mu_\mathrm{B}$ for each data point. 
For a ferromagnetic interchain coupling $J_5<0$, this treatment results in a small region near the saturation in which $M/M_\mathrm{sat}$ is multivalued, as seen in the figure; 
this should be regarded as an artifact of the mean-field treatment. 
}
\label{fig:mag_ex_th}
\end{figure}

The magnetization was calculated using the DMRG and ED methods 
for the $J_1$-$J_2$-$J_5$ spin-chain model with $J_1:J_2:J_5 = -0.42:1:-0.11$, $J_2 = 4.0$ meV, and $g=2.095$. 
The model can conveniently be defined on an approximate rectangular lattice with the primitive vectors $\uv=(b/2,0)$ and $\vv=(0,a)$. 
The $J_1$, $J_2$, and $J_5$ interactions connect spins separated by the vectors $\uv$, $2\uv$, and $\pm \uv+\vv$, respectively. 
The DMRG calculations were performed for the pure 1D $J_1$-$J_2$ model with $N_\mathrm{1D}=168$ spins, and the interchain coupling $J_5$ was taken into account in a mean-field manner \cite{Sta10}.
Namely, at fixed $m=\frac{S^z_{\rm tot}}{N_\mathrm{1D}}$, the ground-state spin-interaction energy per spin, $E_\mathrm{int}^\mathrm{1D}(m)/N_\mathrm{1D}$, is modified by an additive correction of $2J_5 m^2$. Correspondingly, in the magnetization ($m$ versus $h=g\mu_\mathrm{B} H=\frac{1}{N_\mathrm{1D}}\frac{\partial E_\mathrm{int}^\mathrm{1D}(m)}{\partial m}$), $h$ is modified by an additive correction of $4J_5 m$. 
In contrast, the ED calculations were performed directly for the 2D model with 28 spins. 
To reduce finite-size effects in ED, we adopted a screw boundary condition \cite{Nis09}, in which spins separated by $L_x\uv+\vv$ and $L_y\vv$ with $(L_x,L_y)=(7,4)$ were identified.

In Fig.~\ref{fig:mag_ex_th}, the measured magnetization~\cite{Orl17} and the DMRG and ED results are shown together. 
Here, the numerical results are presented for the cases with and without the interchain coupling $J_5$ to highlight its effect. 
We find a good agreement between the experimental data and the numerical results with $J_5$ coupling, 
which supports the present model. 
Furthermore, the agreement between the DMRG and ED results demonstrates the effectiveness of the mean-field treatment of $J_5$ in the DMRG calculations.


\begin{thebibliography}{99}

\bibitem{Kec07}
L. Kecke, T. Momoi, and A. Furusaki, \textit{Phys. Rev. B} \textbf{76}, 060407(R) (2007).

\bibitem{Hik08} 
T. Hikihara, L. Kecke, T. Momoi, and A. Furusaki, \textit{Phys. Rev. B} \textbf{78}, 144404 (2008).

\bibitem{Sud09} 
J. Sudan, A. L\"uscher, and A. M. L\"auchli, \textit{Phys. Rev. B} \textbf{80}, 140402(R) (2009).

\bibitem{Bal16}
L. Balents and O. A. Starykh, \textit{Phys. Rev. Lett.} \textbf{116}, 177201 (2016). 

\bibitem{Lau06} 
A. L\"auchli, G. Schmid, and S. Trebst \textit{Phys. Rev. B} \textbf{74}, 144426 (2006).

\bibitem{Har06} 
K. Harada, N. Kawashima, and M. Troyer \textit{J. Phys. Soc. Jpn.} \textbf{76}, 013703 (2006).

\bibitem{Pro04} 
A. V. Prokofiev, I. G. Vasilyeva, V. N. Ikorskii, V. V. Malakhov, I. P. Asanov, and W. Assmus, \textit{J. Solid State Chem.} \textbf{177}, 3131 (2004).

\bibitem{End05} 
M. Enderle, C. Mukherjee, B. F\r{a}k, R. K. Kremer, J.-M. Broto, H. Rosner, S.-L. Drechsler, J. Richter, J. Malek, A. Prokofiev, W. Assmus, S. Pujol, J.-L. Raggazzoni, H. Rakoto, M. Rheinst\"adter, and H. M. R\o{}nnow, \textit{Europhys. Lett.} \textbf{70}, 237 (2005).

\bibitem{Koo11} 
H.-J. Koo, C. Lee, M.-H. Whangbo, G. J. McIntyre, and R. K. Kremer, \textit{Inorg. Chem.} \textbf{50}, 3582 (2011).

\bibitem{Gib04}
B. J. Gibson, R. K. Kremer, A. V. Prokofiev, W. Assmus, and G. J. McIntyre, \textit{Physica B (Amsterdam)} \textbf{350}, e253 (2004). 

\bibitem{End10}
 M. Enderle, B. F\aa{}k, H.-J. Mikeska, R. K. Kremer, A. Prokofiev, and W. Assmus, \textit{Phys. Rev. Lett.} \textbf{104}, 237207 (2010).

\bibitem{Nai07} 
Y. Naito, K. Sato, Y. Yasui, Y. Kobayashi, Y. Kobayashi, M. Sato, \textit{J. Phys. Soc. Jpn.} \textbf{76}, 023708 (2007). 

\bibitem{Sch08}
F. Schrettle, S. Krohns, P. Lunkenheimer, J. Hemberger, N. B\"uttgen, H.-A. Krug von Nidda, A. V. Prokofiev, and A. Loidl, \textit{Phys. Rev. B} \textbf{77}, 144101 (2008). 

\bibitem{Mou11} 
M. Mourigal, M. Enderle, R. K. Kremer, J. M. Law, and B. F\r{a}k, \textit{Phys. Rev. B} \textbf{83}, 100409(R) (2011).

\bibitem{Ruf19}
A. Ruff, P. Lunkenheimer, H.-A. K. Krug von Nidda, S. Widmann, A. Prokofiev, L. Svistov, A. Loidl, and S. Krohns, \textit{npj Quantum Materials} \textbf{4}, 24 (2019). 

\bibitem{But10}
N. B\"uttgen, W. Kraetschmer, L. E. Svistov, L. A. Prozorova, and A. Prokofiev, \textit{Phys. Rev. B} \textbf{81}, 052403 (2010).

\bibitem{But12}
N. B\"uttgen, P. Kuhns, A. Prokofiev, A. P. Reyes, and L. E. Svistov, \textit{Phys. Rev. B} \textbf{85}, 214421 (2012).

\bibitem{Mas11}
T. Masuda, M. Hagihala, Y. Kondoh, K. Kaneko, and N. Metoki, \textit{J. Phys. Soc. Jpn.} \textbf{80}, 113705 (2011).

\bibitem{Sat13} 
M. Sato, T. Hikihara, and T. Momoi, \textit{Phys. Rev. Lett.} \textbf{110}, 077206 (2013).

\bibitem{Sta14}
O. A. Starykh and L. Balents, \textit{Phys. Rev. B} \textbf{89}, 104407 (2014). 


\bibitem{Mou12} 
M. Mourigal, M. Enderle, B. F\r{a}k, R. K. Kremer, J. M. Law, A. Schneidewind, A. Hiess, and A. Prokofiev, \textit{Phys. Rev. Lett.} \textbf{109}, 027203 (2012).

\bibitem{But14}
N. B\"uttgen, K. Nawa, T. Fujita, M. Hagiwara, P. Kuhns, A. Prokofiev, A. P. Reyes, L. E. Svistov, K. Yoshimura, and M. Takigawa, \textit{Phys. Rev. B} \textbf{90}, 134401 (2014).

\bibitem{Hir19} 
D. Hirobe, M. Sato, M. Hagihala, Y. Shiomi, T. Masuda, and E. Saitoh, \textit{Phys. Rev. Lett.} \textbf{123}, 117202 (2019).

\bibitem{Zhi10} 
M. E. Zhitomirsky and H. Tsunetsugu, \textit{Europhys. Lett.} \textbf{92}, 37001 (2010).

\bibitem{Ued09} 
H. Ueda and K. Totsuka, \textit{Phys. Rev. B} \textbf{80}, 014417 (2009).

\bibitem{Svi11} 
L. E. Svistov, T. Fujita, H. Yamaguchi, S. Kimura, K. Omura, A. Prokofiev, A. I. Smirnov, Z. Honda, and M. Hagiwara , \textit{JETP Lett.} \textbf{93}, 21 (2011).

\bibitem{Orl17} 
A. Orlova, E. L. Green, J. M. Law, D. I. Gorbunov, G. Chanda, S. Kr\"amer, M. Horvati\'c, R. K. Kremer, J. Wosnitza, and G. J. L. A. Rikken, \textit{Phys. Rev. Lett.} \textbf{118}, 247201 (2017).

\bibitem{Gen19} 
M. Gen, T. Nomura, D. I. Gorbunov, S. Yasin, P. T. Cong, C. Dong, Y. Kohama, E. L. Green, J. M. Law, M. S. Henriques, J. Wosnitza, A. A. Zvyagin, V. O. Cheranovskii, R. K. Kremer, and S. Zherlitsyn, 
\textit{Phys. Rev. Research} \textbf{1}, 033065 (2019).

\bibitem{Gra19} 
C. P . Grams, S. Kopatz, D. Br\"uning, S. Biesenkamp, P. Becker, L. Bohat\'y, T. Lorenz, and J. Hemberger, \textit{Sci. Rep.} \textbf{9}, 4391 (2019).

\bibitem{Daou10} 
R. Daou, F. Weickert, M. Nicklas, F. Steglich, A. Haase, and M. Doerr, \textit{Rev. Sci. Instrum.} \textbf{81}, 033909 (2010).

\bibitem{Kru02}H.-A. Krug von Nidda, L. E. Svistov, M. V. Eremin, R. M. Eremina, A. Loidl, V. Kataev, A. Validov, A. Prokofiev, and W. A{\ss}mus, \textit{Phys. Rev. B} \textbf{65}, 134445 (2002).

\bibitem{Sta10} 
O. A. Starykh, H. Katsura, and L. Balents, \textit{Phys. Rev. B} \textbf{82}, 014421 (2010).

\bibitem{titpack} H. Nishimori, TITPACK ver. 2, URL: \url{http://www.qa.iir.titech.ac.jp/~nishimori/titpack2_new/index-e.html} .

\bibitem{Zap08} 
V. S. Zapf, V. F. Correa, P. Sengupta, C. D. Batista, M. Tsukamoto, N. Kawashima, P. Egan, C. Pantea, A. Migliori, J. B. Betts, M. Jaime, and A. Paduan-Filho, \textit{Phys. Rev. B} \textbf{77}, 020404 (2008).

\bibitem{Ike19} 
A. Ikeda, S. Furukawa, O. Janson, Y. H. Matsuda, S. Takeyama, T. Yajima, Z. Hiroi, and H. Ishikawa, \textit{Phys. Rev. B} \textbf{99}, 140412(R) (2019).


\bibitem{Jai12} 
M. Jaime, R. Daou, S. A. Crooker, F. Weickert, A. Uchida, A. E. Feiguin, C. D. Batista, H. A. Dabkowska, and B. D. Gaulin, \textit{Proc. Natl. Acad. Sci. USA} \textbf{109}, 12404 (2012).

\bibitem{Saw05} 
Y. Sawai, S. Kimura, T. Takeuchi, K. Kindo, and H. Tanaka, \textit{Prog. Theor. Phys. Suppl.} \textbf{159}, 208 (2005).

\bibitem{note-pow-fit}
The exponents $p$'s for the correlations on the $J_1$ and $J_2$ bonds were determined from the least-square fitting of the DMRG data with Eq.\ (\ref{eq:corr_scaling}) using the data for $0.4 \le M/M_{\rm sat} \le 0.7$.
On the other hand, the solid line for the correlations on the $J_5$ bond 
simply shows the relation $\overline{\langle{\bf S}_{\textit{i}}\cdot {\bf S}_{\textit{j}}\rangle_{M}}=(M/2M_{\rm sat})^2$ expected in the mean-field treatment. 

\bibitem{Nis09}Y. Nishiyama, \textit{Phys. Rev. B} \textbf{79}, 054425 (2009). 

\bibitem{HMeisnerMK2009}
F. Heidrich-Meisner, I. P. McCulloch, and A. K. Kolezhuk, \textit{Phys. Rev. B}  \textbf{80}, 144417 (2009).

\bibitem{FurukawaSOF2012}
S. Furukawa, M. Sato, S. Onoda, and A. Furusaki, \textit{Phys. Rev. B}  \textbf{86}, 094417 (2012).

\bibitem{HikiharaFL2017}
T. Hikihara, A. Furusaki, and S. Lukyanov, \textit{Phys. Rev. B} \textbf{96}, 134429 (2017).

\bibitem{HikiharaF2004}
T. Hikihara and A. Furusaki, \textit{Phys. Rev. B} \textbf{69}, 064427 (2004).

\bibitem{fitting-quality}
In fact, we found that the quality of the fitting became poor for small $S^z_{\rm tot}$ and near the saturation.
Possible reasons for this include the followings.
Firstly, the prediction of the bimagnon TLL theory is less accurate for small $S^z_{\rm tot}$, where the one-magnon gap tends to be small, and also for $S^z_{\rm tot}$ close to the saturation, 
where the group velocity is small.
Secondly, in deriving the formulas (\ref{eq:cor1-form}) and (\ref{eq:cor2-form}), we neglected the effects of higher-order terms and the possible need to optimize the positions at which the Dirichlet boundary conditions are imposed (see, e.g., Ref.\ \onlinecite{HikiharaFL2017}).
By simply applying the formulas (\ref{eq:cor1-form}) and (\ref{eq:cor2-form}), we found that for those cases of $S^z_{\rm tot}$ the estimates of $c'_n$ and $K$ could not be obtained reliably as they were sensitive to the range of $j$ 
used in fitting the data of 
$\langle {\bf S}_j \cdot {\bf S}_{j+n}\rangle_m^{\rm (o)}$. 
However, even in such cases of $S^z_{\rm tot}$, we could achieve robust estimates of the uniform part $c_n$, which were almost independent of the fitting range and expected to be reliable.

\end{thebibliography}
\end{document}